%% file: main.tex
\documentclass[conference]{IEEEtran}
\IEEEoverridecommandlockouts

\usepackage{cite}
\usepackage{amsmath,amssymb,amsfonts}
\usepackage{graphicx}
\usepackage{textcomp}
\usepackage{xcolor}
\usepackage{booktabs}
\usepackage{multirow}
\usepackage{url}

\usepackage{algorithm} 
\usepackage{algorithmic}

\usepackage{pgfplots}
\pgfplotsset{compat=1.18}

\def\BibTeX{{\rm B\kern-.05em{\sc i\kern-.025em b}\kern-.08em
    T\kern-.1667em\lower.7ex\hbox{E}\kern-.125emX}}

\begin{document}

\title{SeamlessEdit: Background Noise Aware Zero-Shot Speech Editing with in-Context Enhancement}

\author{
    \IEEEauthorblockN{Kuan-Yu Chen\IEEEauthorrefmark{1}\IEEEauthorrefmark{2}, Jeng-Lin Li\IEEEauthorrefmark{2}, De-Yan Lu\IEEEauthorrefmark{1}, and Jian-Jiun Ding\IEEEauthorrefmark{1}}
    
    \IEEEauthorblockA{\IEEEauthorrefmark{1}GICE, National Taiwan University, Taipei, Taiwan}
    \IEEEauthorblockA{\IEEEauthorrefmark{2}AI Research Center, Inventec Corporation, Taipei, Taiwan}
    \IEEEauthorblockA{f13942135@ntu.edu.tw, jenglin.li@inventec.com, qe59979022@gmail.com, jjding@ntu.edu.tw}
}

\maketitle

\begin{abstract}
With the fast development of zero-shot text-to-speech technologies, it is possible to generate high-quality speech signals that are indistinguishable from the real ones. Speech editing, including speech insertion and replacement, appeals to researchers due to its potential applications. However, existing studies only considered clean speech scenarios. In real-world applications, the existence of environmental noise could significantly degrade the quality of generation. In this study, we propose a noise-resilient speech editing framework, SeamlessEdit, for noisy speech editing. SeamlessEdit adopts a frequency-band-aware noise suppression module and an in-content refinement strategy. It can well address the scenario where the frequency bands of voice and background noise are not separated. The proposed SeamlessEdit framework outperforms state-of-the-art approaches in multiple quantitative and qualitative evaluations.
\end{abstract}


\input{sec/1_intro}

\input{sec/2_method}

\input{sec/3_exp_conclusion}
\input{sec/4_conclusion}

\bibliographystyle{IEEEtran} 
\bibliography{mybib}

\end{document}

%% file: sec/1_intro.tex
\section{Introduction}
\label{sec:intro}
Recent advances in zero-shot text-to-speech (TTS) technologies~\cite{AudioLDM2_2024,casanova24_interspeech,eskimez2024e2} have enabled more sophisticated applications beyond straightforward text conversion, including speaker cloning and style transfer using minimal text prompts or acoustic samples. The flexibility of these systems has led to significant innovations in speech editing. It has driven the development of comprehensive applications such as audio post-production, news broadcast correction, and media content revision. The ultimate goal of speech editing research is to seamlessly integrate or modify input speech segments and align the edited output with the source speech without audible quality degradation.

Several techniques can effectively improve the quality of audio generation and eliminate the acoustic discrepancy in the segment boundaries.
Typically, fade-in and fade-out techniques are used to smooth the transitions between cut-and-paste audio segments~\cite{tan2021editspeech}. However, intricate editing procedures may generate artifacts and degrade audio quality. Text prompts are viable for generating the speech signal following a specific style ~\cite{liang20243}. 
P-Flow in ~\cite{kim2023p} uses a prompt-based speaker-conditioned text
representation to improve quality. 
In addition, context-aware methods have been proposed to eliminate unnatural edited boundaries ~\cite{alexos2024attentionstitch}. FluentSpeech~\cite{jiang2023fluentspeech} alleviates the over-smoothing problem using iterative refinement for stutter removal. 
Mapache~\cite{cambara2024mapache} addresses the problem of left-to-right generation in autoregressive (AR) models by a non-AR parallel transformer. FluentEdit~\cite{liu24p_interspeech} applies the acoustic and prosody consistency loss when training the diffusion model for text-based editing. InstructSpeech~\cite{huang_icml2024} is a multitask model and applies a large language model for training. 
 
However, in-the-wild speech editing has often been neglected, leading to unsatisfactory speech editing results in real-world applications. For instance, a speech recorded in a coffee shop with background music is difficult to edit using current methods. Voicebox~\cite{le2023voicebox} has been employed for the in-the-wild setting while the assumption of clean surrounding unedited speech segments may not be always realistic. 
SpeechX~\cite{wang2024speechx} further enhances noisy speech editing through multitask learning. However, the necessity of training still has the risk of over-fitting. 
VoiceCraft~\cite{peng-etal-2024-voicecraft} improves SpeechX by rearranging the neural codec outputs for the bidirectional context and enlarging the token space while still presenting distortion when confronting a variety of background noises. In the field of speech enhancement, research focuses on manipulating noises such as adding background noise to a given speech~\cite{yang2024usee}. 

Modern speech editing methods degrade in noisy environments due to the domain gap between clean training data and noisy test audio. A key challenge is the spectral overlap between speech and background noise, which compromises the generated voice quality and style. To address this, we propose a noise-aware framework that prioritizes preserving clean speech characteristics.

This study presents \textbf{SeamlessEdit}, a noise-robust speech editing framework that introduces in-context refinement for natural integration of speech and background noise. Unlike prior methods that assume clean inputs, SeamlessEdit is designed for real-world acoustic conditions. The framework first separates speech and ambient noise with a score-based diffusion model, then applies sparse Bayesian learning (SBL) filtering to preserve key speech structures while reducing residual noise. The refined signal is incorporated into the decoder for in-context refinement, enabling seamless boundary transitions and enhanced perceptual quality.  

Experiments on the EARS-WHAM dataset~\cite{richter2024ears} demonstrate significant improvements, achieving 29.01\% NMOS and 24.28\% SMOS gains over state-of-the-art systems. Human evaluations further confirm that SeamlessEdit produces audio nearly indistinguishable from natural recordings, effectively blending speech, edits, and ambient noise. Case studies also highlight robustness in challenging scenarios such as overlapping talkers, spontaneous pauses, and dynamic noise. Together, these results establish SeamlessEdit as a practical solution for applications including podcast editing, interview restoration, and archival enhancement.

%% file: sec/2_method.tex
\begin{figure}[htbp]
    \centering
    \includegraphics[width=1\textwidth, height=0.4\textheight, keepaspectratio]{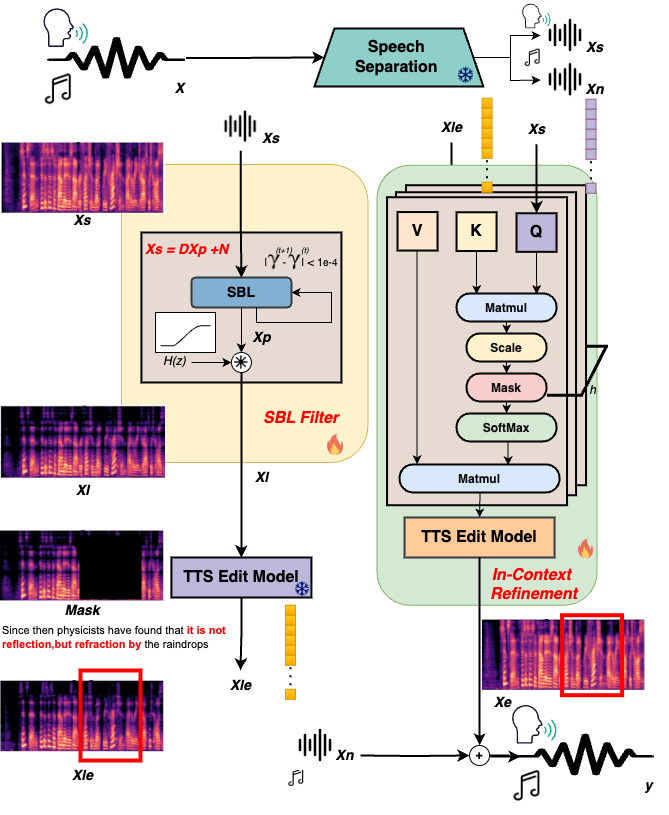} 
    \caption{The proposed SeamlessEdit framework separates human voice $X_s$ and suppresses residual noise to derive an edited speech $X_{le}$. An in-context refinement enhances the editing of $X_s$ using $X_{le}$ for indistinguishable noisy editing results. The SBL filter is also adopted to improve the robustness to noise.}
    \label{fig:model}
\end{figure}

\vspace{-2em}

\section{Methods}

In this study, we aim to address the speech editing problem in noisy scenarios. The proposed SeamlessEdit framework is described in~\ref{ssec:SeamlessEdit} which contains speech separation, residual noise suppression, and in-context refinement.

For our experiments, we used the EARS-WHAM dataset, which combines high-quality 16kHz speech from the EARS corpus~\cite{richter2024ears} with diverse background noises from WHAM!~\cite{Wichern2019WHAM}. The dataset provides expressive and conversational speech mixed with real-world noise at 198 SNR levels. We performed evaluation on both clean and noisy subsets, using 886 speech samples paired with 594 noise recordings, with an average speech duration of 15.6 seconds.

\subsection{SeamlessEdit: Noisy Speech Editing Framework}
\label{ssec:SeamlessEdit}
Given an input speech sequence $X = \{X_1, \ldots, X_T\}$ of length $T$, 
the $m^{\text{th}}$ to $(m+k)^{\text{th}}$ samples specify the target editing region. 
The edited speech is denoted as
\[
X_e = \{X_1, \ldots, X_m, \ldots, X_{m+k}, \ldots, X_T\}.
\]

The proposed SeamlessEdit framework consists of a speech separation module 
(Section~\ref{ssec:speech_separation}) followed by frequency-aware noise suppression 
(Section~\ref{ssec:noise_suppression}), which extracts the separated signal $X_s$, 
the noise-suppressed signal $X_l$, and the background signal $X_n$. 
An in-context refinement strategy further enhances the pre-trained editing model 
(Section~\ref{ssec:backbone}) to generate edited voices under noisy conditions.

\subsection{Speech Separation}
\label{ssec:speech_separation}
The proposed SeamlessEdit framework leverages StoRM~\cite{lemercier2023storm} as a speech separation module to efficiently separate human voice and background noise in a few diffusion steps. This separation module integrates predictive losses with diffusion processes, allowing the reduction of complexity and suppression of artifacts. In contrast to the costly iterative refinement process in other methods, the proposed speech separation model is efficient and robust to distinct acoustic conditions. The speech separation operation can be expressed as $X_s, X_n = \text{StoRM}(X)$ where $X_s$ is the separated speech signal and $X_n$ is the background noise. To achieve a good result for speech separation, $X_s$ should preserve low- and high-frequency speech harmonics even if $X$ is interfered with audible noise patterns.

\begin{table*}[h]
    \centering
    \caption{Insertion and replacement results in noisy conditions using the EARS-WHAM dataset. The superscript ``*'' denotes the clean condition, providing an upper bound of the results. One can see that the proposed SeamlessEdit models achieve the best performance.}
    \renewcommand{\arraystretch}{1.2} 
    \setlength{\tabcolsep}{3pt} 
    \small 
    \resizebox{\textwidth}{!}{
    \begin{tabular}{l c c c c | c c c c | c c c c}
        \toprule
        \multirow{2}{*}{\textbf{Model}} & \multicolumn{4}{c|}{\textbf{Insertion}} & \multicolumn{4}{c|}{\textbf{Short Replacement}} & \multicolumn{4}{c}{\textbf{Long Replacement}} \\
        \cline{2-13}
        & \textbf{WER} $\downarrow$ & \textbf{PES} $\downarrow$ & \textbf{NMOS} $\uparrow$ & \textbf{SMOS} $\uparrow$ 
        & \textbf{WER} $\downarrow$ & \textbf{PES} $\downarrow$ & \textbf{NMOS} $\uparrow$ & \textbf{SMOS} $\uparrow$ 
        & \textbf{WER} $\downarrow$ & \textbf{PES} $\downarrow$ & \textbf{NMOS} $\uparrow$ & \textbf{SMOS} $\uparrow$ \\
        \midrule
        \textbf{Ground Truth} & - & - & - & - & 0.12 & 0.51 & 3.97 & 3.82 & 0.12 & 0.51 & 3.97 & 3.82 \\
        \textcolor{gray}{\textbf{Voicecraft*}} & \textcolor{gray}{0.18} & \textcolor{gray}{0.56} & \textcolor{gray}{4.23} & \textcolor{gray}{4.11} 
        & \textcolor{gray}{0.10} & \textcolor{gray}{0.52} & \textcolor{gray}{4.25} & \textcolor{gray}{4.19} 
        & \textcolor{gray}{0.10} & \textcolor{gray}{0.54} & \textcolor{gray}{4.32} & \textcolor{gray}{4.23} \\
        \midrule
        \textbf{FluentSpeech} & - & - & - & - & 0.23 & 1.82 & 1.67 & 1.66 & 0.22 & 1.84 & 1.50 & 1.49 \\
        \textbf{Voicecraft} & 0.33 & 1.28 & 2.93 & 2.76 & 0.24 & 1.23 & 2.9 & 2.99 & 0.23 & 1.16 & 2.93 & 2.60 \\
        \textbf{Our SeamlessEdit w/o ICL} & \textbf{0.28} & 0.89 & 3.75 & \textbf{3.48} 
        & 0.23 & 0.80 & 3.38 & 3.12 & 0.24 & 0.86 & 3.42 & \textbf{3.13} \\
        \textbf{Our SeamlessEdit} & \textbf{0.28} & \textbf{0.77} & \textbf{3.78} & 3.43 
        & \textbf{0.22} & \textbf{0.72} & \textbf{3.56} & \textbf{3.13} 
        & \textbf{0.22} & \textbf{0.75} & \textbf{3.65} & 3.11 \\
        \bottomrule
    \end{tabular}
} 
    \label{tab:performance_comparison} 
\end{table*}

\subsection{Frequency-Band Aware Noise Suppression}
\label{ssec:noise_suppression}
Speech signal separation is challenging if its support in the frequency domain is intertwined with that of the noise, which can bias the posterior estimation result. Therefore, we propose a noise suppression module to strengthen speech intelligibility while attenuating residual noise for unbiased voice representation. 

We employ sparse Bayesian learning (SBL)~\cite{10448419,xenaki2018sound} to estimate the sparse frequency bands of the human voice and thus eliminate residual noise whose spectrum is usually uniformly distributed. SBL assumes that signals exhibit sparsity in an appropriate transform domain. 

Specifically, $X_s$ is decomposed into the human voice $X_p$ and residual noise $N$ as follows: 

 \begin{equation}
 X_s = D X_p + N,
 \end{equation} 
where $D$ is an over-complete dictionary of basis and $X_p$ is constrained to be sparse in the frequency domain. The obtained $X_p$ is refined by iterative posterior estimation that maximizes the evidence function. Given an initialized coefficient vector $\mu^{(t)}$, the parameters are calculated as follows in each iteration:

\begin{equation}
\gamma^{(t+1)} = \frac{1}{\left| \mu^{(t)} \right| + \epsilon},
\end{equation}

\begin{equation}
\Sigma^{(t+1)} = \left( \frac{D^T D}{\lambda} + \text{diag}(1 / \gamma^{(t)}) \right)^{-1}, 
\end{equation}

\begin{equation}
\mu^{(t+1)} = \Sigma^{(t+1)} D^T X_s / \lambda,
\end{equation}
where $\gamma_i$ controls sparsity, $\lambda$ is a regularization parameter to balance the sparsity and the reconstruction accuracy, and $\epsilon$ prevents the division from being zero.
The iteration ends when $\gamma$ converges beyond a small threshold $\delta$: $| \gamma^{(t+1)} - \gamma^{(t)} | < \delta$.

Once the sparse coefficients $X_p$ are estimated, the speech signal undergoes a Butterworth filter to enhance the desired frequency components while suppressing residual noise. The transfer function of the Butterworth filter is defined as follows:

\begin{equation}
H(z) = \frac{\sum_{k=0}^{M} b_k z^{-k}}{1 + \sum_{k=1}^{N} a_k z^{-k}},
\end{equation}
where $b_k$ and $a_k$ are the filter coefficients in the numerator and the denominator, respectively. To minimize phase distortion, forward and backward filtering are applied on $X_p$ using the SBL-enhanced filter $H(z)$, ensuring that the final speech signal $X_l$ maintains high intelligibility:

\begin{equation}
X_l = \text{SBL Filter}(\delta, b, a, X_s).
\end{equation}

\subsection{Neural Codec Editing Model for TTS}
\label{ssec:backbone}
We adopt VoiceCraft~\cite{peng-etal-2024-voicecraft} as the neural codec editing model for TTS to perform insertion, deletion, and substitution flexibly and efficiently. Its causal masking and delayed stacking mechanisms enhance the codec backbone for discrete tokenization while simplifying diffusion steps.

For noisy speech editing, we use a separated speech input $X_s=\{X_{s,1}, …, X_{s,T}\}$ and the target region mask $M=\{m, …, m+k\}$ within the neural codec editing model $f$ to generate edited speech signals $X_e=\{X_{e,1}, …, X_{e,T}\}$. The edited region $\{X_{e,m}, …, X_{e,m+k}\}$ maintains quality consistent with non-edited regions, ensuring seamless integration. VoiceCraft is adopted to perform autoregressive sequence prediction conditioned on the speech transcript.

To further enhance speech quality, we apply SBL filters to suppress residual noise and refine the spectral characteristics of separated speech signals. The filtered outputs $X_l = \{X_{l,1}, …, X_{l,T}\}$ are served as the inputs of the TTS editing model $g$ to generate edited speech signals $X_{le} = \{X_{le,1}, …, X_{le,T}\}$. Given a masked target region $M_l$, the TTS editing model modifies this segment while preserving the surrounding regions. The autoregressive generation process, conditioned on linguistic and acoustic representations, ensures natural and coherent synthesis.

\begin{figure*}[t]
    \centering
    \includegraphics[width=\textwidth,height=0.2\textheight]{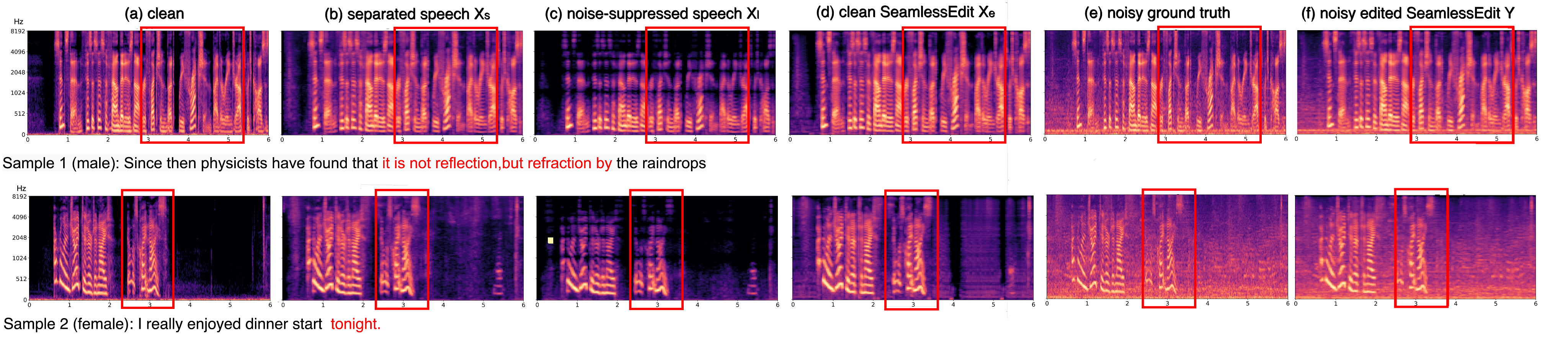} 
    \vspace{-7mm}
    \caption{
    Mel-spectrograms of each stage of the proposed model; (a) the clean signal; (e) the noisy ground truth. Different editing stages include (b) separated speech $X_s$, (c) noise-suppressed speech $X_{l}$, (d) SeamlessEdit processed clean speech $X_e$, and (f) the final noisy editing result $Y$ of the proposed SeamlessEdit model. Red texts and boxes denote the edited regions.}
   
    \vspace{-3mm}
    \label{fig:mel}
\end{figure*}

\subsection{In-Context Refinement}
\label{ssec:in_context_refinement}
In $\S$\ref{ssec:backbone}, the auditory quality of the generated speech signal from the neural codec editing model $f$ is naturally limited by its intricate conjunction between the human voice and background noise. Inspired by the success of prompt design~\cite{suh24_interspeech,chang24c_interspeech} and in-context learning~\cite{chang22e_interspeech} in speech recognition, we introduce an in-context learning strategy to refine the editing process by integrating the low-frequency speech $X_{le}$ described in $\S$\ref{ssec:backbone} to emphasize the frequency components of human voices. 
Note that previous in-context learning studies for speech generation usually enhance descriptive attributes such as timbre or spoken content~\cite{Mega-TTS2024,Seed-tts} without incorporating additional acoustic information to refine the speech quality. 

The proposed in-context refinement strategy is implemented by a \textit{MultiHead} attention module before the TTS model $f$ that applies low-frequency embedding $X_{le}$ to enhance $X_s$ in $\S$\ref{ssec:speech_separation} with intrinsic human voice characteristics. In Figure \ref{fig:model}, the input of $Q$ is embedded with $X_s$ while the inputs of $K$ and $V$ are embedded with the low-frequency speech signals $X_l$. $Q$, $K$, and $V$ embeddings are derived by projecting with learnable projection matrices $W_i^QX_s$, $W_i^KX_l$, and $W_i^VX_l$, respectively. 
Then, the attention of the $i^{th}$ head is determined as follows:
\begin{equation}
\text{head}_i = \text{softmax} \left( \frac{Q K^T}{\sqrt{d_k}} \right) V, 
\end{equation}
where $d_k$ is the dimension of $K$ embedding. Concatenating $h$ attention heads, the output $X_{e}$ of the MultiHead module is:
\begin{equation}
X_{e} = \text{MultiHead}(Q, K, V) = [\text{head}_1, \dots, \text{head}_h] W^O,
\end{equation}
where $W^O$ is the weight. 
After in-context refinement, the separated speech $X_s$ is close to the source speech signal. The final edited speech $X_e$ encompasses the detailed characteristic of the human voice preserved in $X_{le}$ and maintains auditory consistency at the boundaries. We add the separated noise $X_n$ back to $X_{e}$ and obtain the final noisy editing output $Y$, i.e., $Y=X_{e}+X_n$.

%% file: sec/3_exp_conclusion.tex
\section{Experiments}

\begin{figure}[htbp]
    \centering
    \includegraphics[width=1\columnwidth]{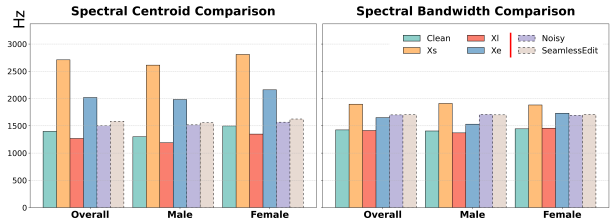}
    \caption{Spectral centroid and bandwidth statistics for short replacement results. We present the result of each editing step in Figure~\ref{fig:mel} for clean and noisy conditions. An ideal editing result should be indistinguishable from the noisy ground.} 
    \label{fig:spectral}
\end{figure}

\subsection{Evaluation Setup}
Experiments are conducted on the EARS-WHAM dataset, containing 886 speech samples and 594 noise recordings across 198 SNR levels. We evaluate both noisy and clean subsets, using the latter to establish an upper-bound reference. The editing tasks include insertion, short replacement (1--6 words), and long replacement (7--12 words). 

The compared baselines include FluentSpeech and Voicecraft. Although FluentSpeech cannot perform insertion, we regard it for short and long replacement tasks. We also apply the ground truth and Voicecraft's clean condition results as the upper-bound reference.

We adopt the evaluation metrics of the word error rate (WER), the perceptual edited score (PES), the naturalness mean opinion score (NMOS), and the subjective mean opinion score (SMOS). The WER measures transcription intelligibility by the Whisper medium~\cite{Whisper}. The PES (range [0, 2]) represents the number of artificial boundaries (0, 1, or 2) detected by listeners. The NMOS assesses relative speech naturalness through pairwise comparisons of fluency and coherence between edited and unedited regions. The SMOS evaluates absolute quality based on human perception.

We recruited 30 subjects to perform subjective evaluation for the NMOS and SMOS (range [1, 5]). To ensure consistency, the same group of evaluators assessed all systems, including baselines. A lower WER and PES, and a higher NMOS and SMOS, mean better performance.

\noindent\textbf{Implementation Details:} We use a single NVIDIA A16 GPU. The TTS model ($\S$\ref{ssec:backbone}) follows the VoiceCraft configuration, using a 50kHz tokenizer with a vocabulary size of 2048. The architecture consists of 16 transformer layers (2048 hidden, 8192 output nodes) followed by 4 dense layers. The noise suppression coefficients ($\S$\ref{ssec:noise_suppression}) are $a=[ 1, 0, 0.486029, 0, 0.017665]$ and $b=[ 0.093981, -0.375923, 0.563885, -0.375923, 0.093981]$.

\subsection{Main Results}
As shown in Table~\ref{tab:performance_comparison}, the proposed SeamlessEdit model achieves $3.78$, $3.56$, and $3.65$ NMOS in insertion, short, and long replacement tasks, respectively. The SMOS results of SeamlessEdit also outperform those of other methods. The performance of the existing method is significantly degraded in noisy scenarios. The state-of-the-art model, VoiceCraft, shows $-1.30$, $-1.35$, and $-1.39$ NMOS differences in insertion, short, and long replacement. Similarly, the SMOS dropped by an average of 1.39 for the three tasks. This is because noise hinders the neural codec model from capturing real human voice characteristics. FluentSpeech exhibits a comparable WER to VoiceCraft but a lower MOS in replacement tasks due to its direct segment replacement, which disrupts waveform continuity. 

Moreover, the proposed SeamlessEdit model attains $0.72$ PES, demonstrating its smoother and more natural editing results. 
We examine the results of SeamlessEdit w/o in-context learning (ICL) to illuminate the effects of in-context refinement. As a result, SeamlessEdit w/o ICL obtains a comparable SMOS, a lower NMOS, and a higher PES compared to SeamlessEdit. The results verify that the refinement mechanism can suppress the noise effect and improve the speech editing performance. 

\subsection{Spectrogram Analysis}
Fig.~\ref{fig:mel} and Fig.~\ref{fig:spectral} show the visual comparison of Mel spectrograms and quantitative spectral characteristics, highlighting the effects of different processing steps. 
In Fig.~\ref{fig:mel}, we compare the edited speech in each stage with the clean ground truth in Fig.~\ref{fig:mel} (a).
The separated speech $X_s$ in Fig.~\ref{fig:mel} (b) shows indistinct and blurry boundaries around the voiced segments owing to residual noise. The SBL-filtered spectrogram in Fig.~\ref{fig:mel} (c) has suppressed more noise, but excessive filtering can over-attenuate high-frequency components, leading to a muffled sound. 
SeamlessEdit in Fig.~\ref{fig:mel} (d) balances the goals of noise removal and the preservation of high-frequency voice component via ICL refinement. 
Therefore, the final noisy editing results in Fig.~\ref{fig:mel} (f) are indistinguishable from the noisy ground truth in Fig.~\ref{fig:mel} (e). 
There is a challenge of distinctly isolating female voices from background noise, as Sample 2 shows. The residual frequency components still appear in the spectrogram despite applying separation and noise suppression techniques.

We used the spectral centroid and the bandwidth~\cite{4412892,4960678} to measure the intensity-weighted average frequency and the average frequency band disparity in Fig.~\ref{fig:spectral}. These two metrics have higher values for the separated speech signal $X_s$ and lower values for the noise-suppressed signal $X_l$. The spectral bandwidth of $X_l$ stays close to that of the clean speech signal while the centroid values are lower. The results are consistent with Fig.~\ref{fig:mel} that $X_l$ invades parts of high-frequency voices and $X_s$ remains intertwined with noise. The difficulty of resolving the overlapped high-frequency bands manifests in female speakers. In Fig.~\ref{fig:spectral}, the female voices are more susceptible to noise interference. Similar evidence is shown in the spectral bandwidth of $X_e$ that the average value of females is higher than that of males. 
Intriguingly, increased timbre preservation can be achieved by the proposed ICL refinement mechanism that leverages the synthesis of the noise-suppressed signal as a reference example.

%% file: sec/4_conclusion.tex
\section{Conclusion}
This work presents \textbf{SeamlessEdit}\footnote{\url{https://danielchen1128.github.io/SeamlessEdit/}}, a noise-robust speech editing framework designed for real-world conditions. Unlike prior methods limited to clean studio recordings, SeamlessEdit can handle diverse background noises while preserving useful ambience, ensuring intelligibility and naturalness of edited speech. By jointly modeling content, prosody, and environmental acoustics, it enables seamless word- and phrase-level editing without audible artifacts. Applications include podcast production, interview restoration, and archival enhancement, highlighting its practicality in everyday and professional audio workflows.